\documentclass[prb,twocolumn,amsmath]{revtex4-1}
\usepackage{graphicx}
\usepackage{dcolumn}
\usepackage{bm}
\usepackage[breaklinks]{hyperref}
\usepackage[all]{hypcap}
\usepackage{color}
\usepackage{array}

\begin{document}


\title{Tunnel diode oscillator measurements of the upper critical magnetic field of FeTe$_{0.5}$Se$_{0.5}$}

\author{Alain Audouard}
\email[E-mail address: ]{alain.audouard@lncmi.cnrs.fr}
\affiliation{Laboratoire National des Champs Magn\'{e}tiques
Intenses (UPR 3228 CNRS, INSA, UJF, UPS) 143 avenue de Rangueil,
F-31400 Toulouse, France}

\author{Lo\"{\i}c Drigo}
\email[E-mail address: ]{loic.drigo@lncmi.cnrs.fr}
\affiliation{Laboratoire National des Champs Magn\'{e}tiques
Intenses (UPR 3228 CNRS, INSA, UJF, UPS) 143 avenue de Rangueil,
F-31400 Toulouse, France}

\author{Fabienne Duc}
\email[E-mail address: ]{fabienne.duc@lncmi.cnrs.fr}
\affiliation{Laboratoire National des Champs Magn\'{e}tiques
Intenses (UPR 3228 CNRS, INSA, UJF, UPS) 143 avenue de Rangueil,
F-31400 Toulouse, France}

\author{Xavier~Fabr\`{e}ges}
\altaffiliation{Present address: Laboratoire L\'{e}on Brillouin, CEA-CNRS UMR 12, 91191 Gif-sur-Yvette Cedex, France.}
\affiliation{Laboratoire National des Champs Magn\'{e}tiques
Intenses (UPR 3228 CNRS, INSA, UJF, UPS) 143 avenue de Rangueil,
F-31400 Toulouse, France}

\author{Ludovic Bosseaux}
\affiliation{Laboratoire National des Champs Magn\'{e}tiques
Intenses (UPR 3228 CNRS, INSA, UJF, UPS) 143 avenue de Rangueil,
F-31400 Toulouse, France}

\author{Pierre Toulemonde}
\affiliation{Univ. Grenoble Alpes, Institut N\'{E}EL, F-38042 Grenoble, France.\\ CNRS, Institut N\'{E}EL, F-38042 Grenoble, France.
}%

\date{\today}


%

\begin{abstract}
Temperature dependence of the upper critical magnetic field ($H_{c2}$) of single crystalline FeTe$_{0.5}$Se$_{0.5}$ ($T_c$ = 14.5 K) have been determined by tunnel diode oscillator-based measurements in magnetic fields of up to 55 T and temperatures down to 1.6 K. The  Werthamer-Helfand-Hohenberg model accounts for the data for magnetic field applied both parallel (\textbf{H} $\|$ $ab$) and perpendicular (\textbf{H} $\|$ $c$) to the iron conducting plane, in line with a single band superconductivity. Whereas Pauli pair breaking is negligible for \textbf{H} $\|$ $c$, Pauli contribution is evidenced for \textbf{H} $\|$ $ab$ with Maki parameter $\alpha$ = 1.4, corresponding to Pauli field $H_{P}$ = 79 T. As a result, the $H_{c2}$ anisotropy ($\gamma$ = $H_{c2}^{ab}$/$H_{c2}^{c}$) which is already rather small at $T_c$ ($\gamma$ = 1.6) further decreases as the temperature decreases and becomes smaller than 1 at liquid helium temperatures.
\end{abstract}


\maketitle

\section{Introduction}

The discovery of superconductivity in La(O$_{1-x}$F$_x$)FeAs with $T_c$ as high as 26 K\cite{Ka08} and, few months later, in Sm(O$_{1-x}$F$_x$)FeAs with $T_c$ = 55 K\cite{Re08} (both of them illustrating the '1111' pnictide family)  raised a tremendous interest for iron-based superconductors, notably in connection with the study of the interplay between superconductivity and magnetism (for recent reviews, see $e.g.$ \cite{Zh11,Le12,Co13}). In that respect, determination of the temperature dependence of the upper critical magnetic field ($H_{c2}$) and its anisotropy yields information on the pair breaking mechanism and allows to decide between single and multigap superconductivity. Regarding the superconducting gap topology, the issue remains under debate since, for example, a nodeless two-gap picture is in agreement with the ARPES data of Ba$_{0.6}$K$_{0.4}$Fe$_2$As$_2$\cite{Di08,Zh08} whereas nodal gaps have been inferred for BaFe$_2$(As$_{1-x}$P$_x$)$_2$\cite{Zh12,Yo13}, these two compounds belonging to the same '122' pnictide family.

Even though large magnetic fields are required to explore the low temperature part of the phase diagrams due to $H_{c2}$ values as high as several tens of a Tesla, numerous works have been devoted to this issue. Reported data share a common feature, namely the decrease of the $H_{c2}$ anisotropy as the temperature decreases, which nevertheless still requires a clear understanding. Oppositely, puzzling results regarding the temperature dependence of $H_{c2}$ can be found in the literature. Indeed, in several cases, (i) the temperature dependence of $H_{c2}$ exhibit a negative curvature consistent with the  Werthamer-Helfand-Hohenberg (WHH) model, eventually including a Pauli contribution, for both \textbf{H} parallel and perpendicular to the conducting $ab$ plane  (\textbf{H} $\|$ $ab$ and \textbf{H} $\|$ $c$, respectively). However, in several other cases, (ii) although the WHH model accounts for the data for \textbf{H} $\|$ $ab$,  either a two-gap behaviour or a roughly linear temperature dependence is observed for \textbf{H} $\|$  $c$.  In this latter case, even though multiple gap superconductivity is in agreement with ARPES data\cite{Di08,Zh08}, it remains to explain why a two-gap behaviour is observed for one of the considered magnetic field directions, only. Alternatively, field-dependent spin flip has been invoked to account for this behaviour\cite{Ko13}.  Besides, for few compounds, (iii) $H_{c2}$ exhibits an upward curvature for the two field directions. More specifically, limiting ourselves to the Fe$_{1+\delta}$Te$_x$Se$_{1-x}$  chalcogenide compounds (referred to as the '11' family), all the above mentioned behaviours have been observed. Namely, behaviours (i)\cite{Br10,Kh10,Kl10,Le10,Se10,Ve13}, (ii)\cite{Se10,Fa10} and (iii)\cite{Be10,Le11} have been reported. An important question deals with the probe used for the determination of $H_{c2}$. As an example, while either behaviour (i) or (iii) is reported for magnetic torque data (yielding the irreversibility field), behaviour (i) is observed in specific heat data\cite{Se10,Kl10}. Nevertheless, discrepancies are still observed using the same probe. As an example, specific heat data collected close to $T_c$ can yield strongly different anisotropy ratios ranging from $\gamma \sim$ 4\cite{Kl10} to nearly isotropic behaviour\cite{Br10}. In that respect, crystal stoichiometry, doping and microstructure might influence the superconducting properties\cite{Ts11}.

Tunnel diode oscillator (TDO) based technique, which is known to be sensitive to the London magnetic penetration depth\cite{Pr11} has already been successfully used to probe $H_{c2}$ relevant to compounds belonging to the '122' pnictide family\cite{Mu11,Ga11,Ro12,Ga13}. Indeed, this contactless technique is very sensitive to superconducting transitions, yielding very high signal-to-noise ratio. Despite of that, to our best knowledge, no data obtained with this technique have been reported for the '11' chalcogenide superconductors, yet. Therefore, this paper reports on the determination, through TDO-based measurements, of the temperature dependence of the upper critical field of a '11' superconductor, namely FeTe$_{0.5}$Se$_{0.5}$, which has already been studied by magnetic torque, resistivity and specific heat\cite{Be10,Se10,Kl10} measurements. The deduced physical parameters (zero-field superconducting temperature, coherence length, orbital and Pauli fields) are compared to the data obtained from these previous measurements.

\section{Experimental}

The studied Fe$_{1+\delta}$Te$_{0.5}$Se$_{0.5}$ ($\delta$ = 0.05) single crystal has been synthesized using the sealed quartz tube method, as detailed in Ref.~\onlinecite{Kl10}.

As reported in Ref.~\onlinecite{Dr10}, the device for radio frequency measurements is a LC-tank circuit powered by a
tunnel diode oscillator (TDO) biased in the negative resistance
region of the current-voltage characteristic. This device is connected to a pair of compensated coils made with copper wire (40 $\mu$m in
diameter). Each of these coils is wound around a Kapton tube of 1.5 mm in diameter. The studied crystal, which is a platelet with dimensions of roughly 1.4 $\times$ 1.4 $\times$ 0.04 mm$^3$, is placed at the centre of one of them with the normal to the conducting $ab$ plane parallel to the coil axis. The fundamental resonant frequency $f_0$ of the whole set is in the range
16 to 20 MHz. This signal is amplified, mixed with a frequency $f$ about 1 MHz below the fundamental frequency and demodulated. Resultant frequency $\Delta f$ = $f - f_0$ has been measured in the temperature range from 1.6 K to
16 K, both in zero-field and in pulsed magnetic fields of up to 55 T with a pulse decay duration of 0.32 s. Magnetic field direction, either parallel or perpendicular to the $ab$ plane, was explored by tilting the compensated coils so that the excitation field was always perpendicular to the $ab$ plane.  It has been checked that the data collected during the raising and the falling part of the pulse coincide, indicating that  no discernible temperature change occurred during the field sweep.

\section{Results and discusion}

\begin{figure} 
\centering
\includegraphics[width=0.9\columnwidth,clip,angle=0]{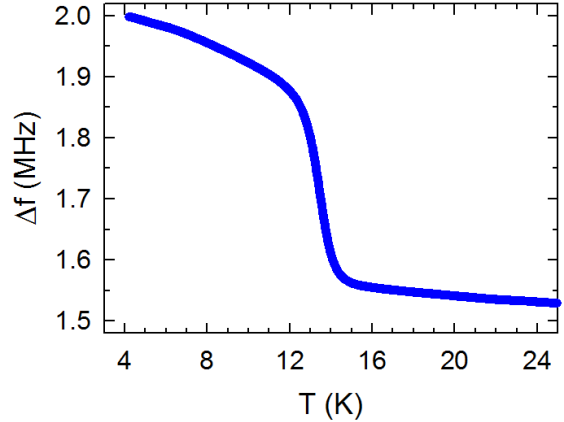}
\caption{\label{fig:ft}(color on line) Zero-field temperature dependence of the TDO frequency. }
\end{figure}

\begin{figure} 
\includegraphics[width=0.9\columnwidth,clip,angle=0]{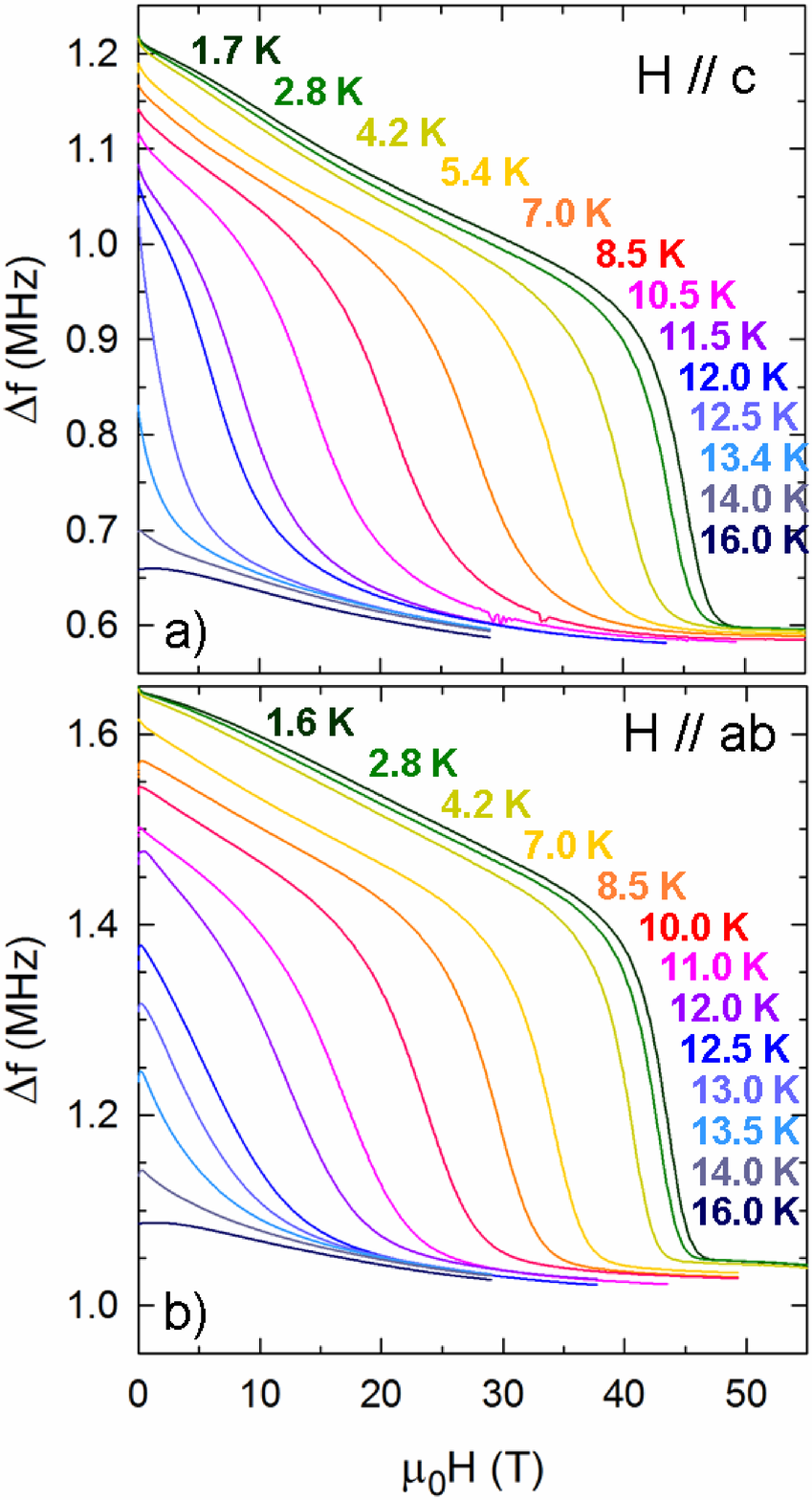}
\caption{\label{fig:fB}(color on line) Field dependence of the TDO frequency at various temperatures for magnetic field applied parallel to (a) the $c$ direction and (b) the $ab$ plane. }
\end{figure}

\begin{figure}                                                    
\centering
\includegraphics[width=0.9\columnwidth,clip,angle=0]{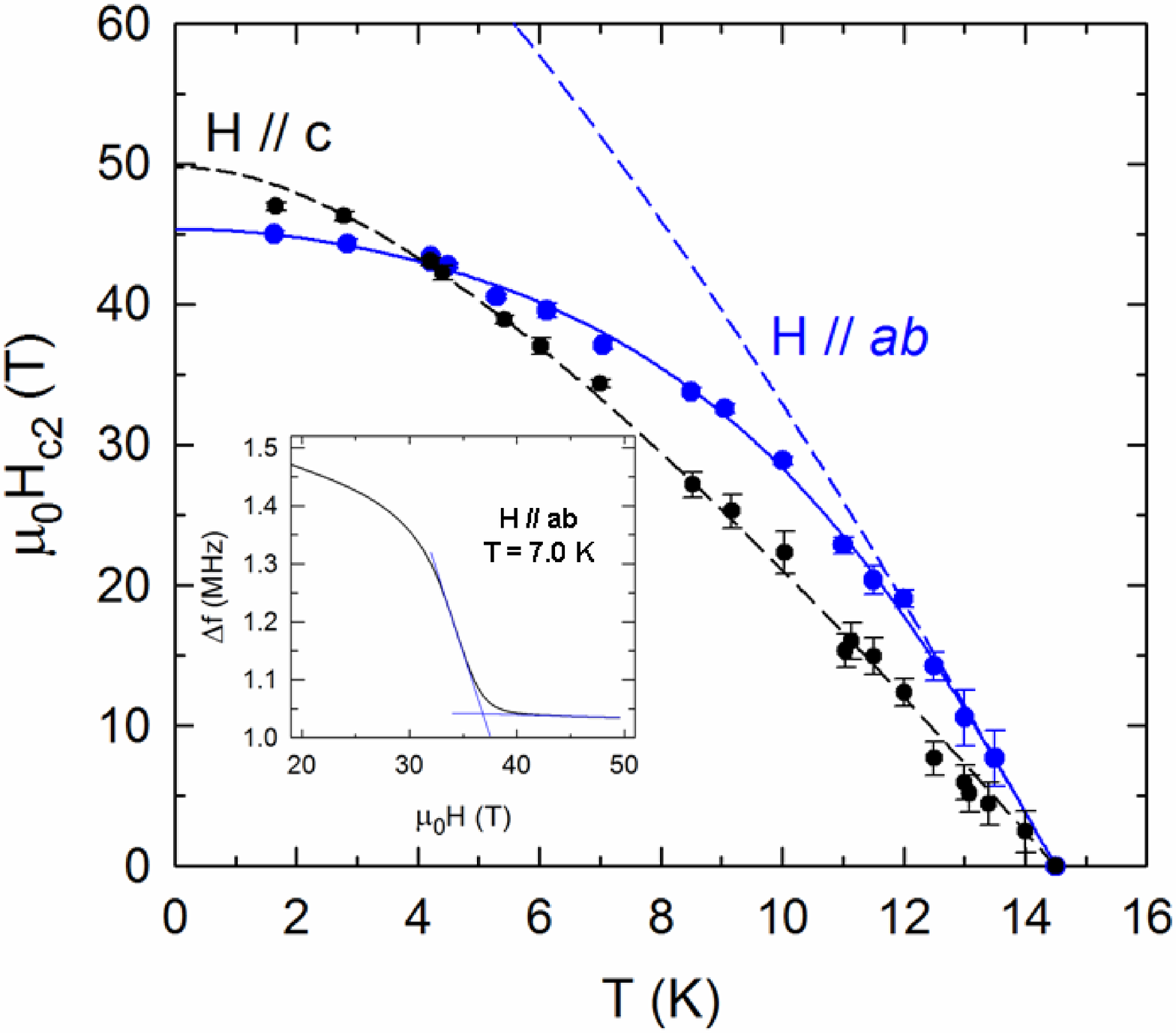}
\caption{\label{fig:Hc2}(color on line) Temperature dependence of the upper critical field $H_{c2}$ for magnetic field applied parallel to either the $ab$ plane (blue symbols) or the $c$ direction (black symbols). Dashed lines are the best fits of the WHH model to either the high temperature part of the data (for \textbf{H} $\|$ $ab$) or down to the lowest temperature (for \textbf{H} $\|$ $c$). Solid line is the best fit to the data for \textbf{H} $\|$ $ab$, including a Pauli contribution ($\mu_0H_{P}$ = 79 T). $H_{c2}$ values are determined according to the construction lines exemplified in the insert. }
\end{figure}

The zero-field TDO frequency displayed in Fig.~\ref{fig:ft}  evidences a large increase as the temperature decreases. This feature can be ascribed to the decrease of the magnetic London penetration depth due to the superconducting transition\cite{Pr11}. As reported in the case of '122' pnictides\cite{Ga11},  the onset of the frequency rise coincides with the superconducting transition deduced from $e.g.$ magnetization or resistivity data. In line with this statement, the measured value ($T_c$ = 14.5 K) is in agreement with specific heat data\cite{Kl10} and corresponds to the best quality samples of the considered composition\cite{Ts11}.

Field-dependent TDO frequency is displayed in Figs.~\ref{fig:fB}(a) and (b) for \textbf{H} $\|$ $c$ and \textbf{H} $\|$ $ab$, respectively. Well defined transitions are observed allowing to determine the temperature dependence of the upper critical magnetic field $H_{c2}$ for the two considered field directions according to the method displayed in the insert of Fig.~\ref{fig:Hc2}. First, in contrast with irreversibility field data\cite{Be10}, a negative curvature is observed in both cases as reported in Fig.~\ref{fig:Hc2}. Such a behaviour, referred above to as 'behaviour (i)', is in qualitative agreement with specific heat\cite{Kl10,Se10} and resistivity\cite{Br10,Kh10,Le10,Ve13} measurements and indicates a single band superconductivity, in line with ARPES measurements\cite{Mi12}. These data yield  $dH_{c2}^{ab}/dT\mid_{T=T_c}$ = -4.9 T/K and  $dH_{c2}^{c}/dT\mid_{T=T_c}$ = -7.7 T/K, hence $\gamma$ = 1.6.  As previously stated\cite{Co13}, the slopes deduced from specific heat data are generally higher than those deduced from magnetic torque and resistivity data. Actually, the values deduced from Fig.~\ref{fig:Hc2}, which are significantly lower than those deduced from specific heat measurements for single crystals belonging to the same batch ($dH_{c2}^{ab}/dT\mid_{T=T_c}$ = -12 T/K and  $dH_{c2}^{c}/dT\mid_{T=T_c}$ = -45 T/K, respectively\cite{Kl10}) are in agreement with resistivity\cite{Br10} and magnetic torque\cite{Se10} data. This result suggests that other contributions such as in-plane resistivity variations and vortices dynamics enter the TDO data around $T_c$. The above mentioned anisotropy value is rather small owing to the two-dimensional character of the Fermi surface. Nevertheless, it stays within the very scattered range reported in the literature. Indeed, limiting ourselves to specific heat data, values as different as $\gamma$ = 1.2\cite{Se10} and $\gamma$ = 4\cite{Kl10} have been reported.

Dashed lines in Fig.~\ref{fig:Hc2} are the best fits of the WHH model\cite{WHH} to the data, assuming negligible spin-orbit coupling\cite{Zh11}. In this framework, the temperature-dependent upper critical field is given by ln(1/t) = $\psi(1/2+h/2t)-\psi(1/2)$ where $\psi$ is the digamma function, $t=T/T_c$ and $h=4\mu_0H_{c2}/[\pi^2(-dH_{c2}/dt)\mid_{t=1}]$. Orbital fields deduced from the data in Fig.~\ref{fig:Hc2}  ($\mu_0H_{c2}^c(0)$ = -0.69$T_c dH_{c2}/dT\mid_{T=T_c})$ are $\mu_0H_{c2}^c(0)$ = 49 T and   $\mu_0H_{c2}^{ab}(0)$ = 78 T. These values are close to those deduced from resistivity measurements ($\mu_0H_{c2}^c(0)$ = 48 T and   $\mu_0H_{c2}^{ab}(0)$ = 63 T)\cite{Br10}. However, larger values are deduced from specific heat data ($\mu_0H_{c2}^c(0)$ = 130 T and   $\mu_0H_{c2}^{ab}(0)$ = 400 T\cite{Kl10}). Hence, the deduced coherence lengths ($\xi_{c}$ =$\sqrt{(\phi_0 H_{c2}^c(0))/(2\pi)}/H_{c2}^{ab}(0)$ =  1.7 nm and $\xi_{ab}$ =$\sqrt{\phi_0/2\pi H_{c2}^c(0)}$ = 2.6 nm) are closer to that derived from resistivity measurements than from specific heat data ($\xi_{c}$ =  0.4 nm and $\xi_{ab}$  = 1.5 nm)\cite{Kl10}.

While this model reproduces fairly well the temperature dependence of $H_{c2}$ for \textbf{H} $\|$ $c$, Pauli pair breaking contribution must be included for \textbf{H} $\|$ $ab$. In this case, the orbital critical field is reduced as $\mu_0H_P=\mu_0H_{c2}^{orb}/\sqrt{1+\alpha^2}$ where the Maki parameter is given by  $\alpha = \sqrt{2}H_{c2}^{orb}/H_P$. A very good agreement with experimental data is obtained with a Pauli field $\mu_0H_P$ = 79 T, $i.e.$ $\alpha$ = 1.4 (see solid line in Fig.~\ref{fig:Hc2}) yielding $\mu_0H_{c2}(0)$ = 45 T. Whereas no reliable $\mu_0H_P$ value could be derived from specific heat data obtained at temperatures above $\sim$ 10 K, in fields below 28 T\cite{Kl10}, the measured value is close to that deduced from resistivity measurements data ($\mu_0H_P$ = 69 T)\cite{Br10}. It should also be noted that the data for \textbf{H} $\|$ $c$ can be accounted for by including a small Pauli contribution ($\alpha$ = 0.25), still in agreement with resistivity data. In addition, assuming strong spin-orbit coupling would yield even larger $\alpha$ values \cite{Or79}. Therefore, Pauli contribution for both field directions cannot be excluded.

Finally, a striking result is the anisotropy inversion occurring at 4 K ($\gamma$ = 0.9 at  1.6 K). This feature, which is in line with the widely observed reduction of the anisotropy as the temperature decreases\cite{Zh11},  has already been observed at 4 K in magnetic torque\cite{Se10} and resistivity\cite{Br10} measurements of FeTe$_{0.5}$Se$_{0.5}$. Such anisotropy inversion which had also been inferred from the extrapolation towards low temperature of specific heat data, albeit at higher temperature ($\sim$ 9 K)\cite{Kl10} remains to be understood.

\section{Summary and conclusion}

Magnetic field- and temperature-dependent superconducting transition of single crystalline FeTe$_{0.5}$Se$_{0.5}$ have been studied by contactless tunnel diode oscillator-based measurements. In zero-field, the temperature dependence of the TDO frequency yields a superconducting transition temperature $T_c$ = 14.5 K which is in agreement with previous specific heat, magnetic torque and resistivity data.

The WHH model accounts for the temperature dependence of the upper critical magnetic field for magnetic field applied both parallel (\textbf{H} $\|$ $ab$) and perpendicular to the conducting $ab$ plane (\textbf{H} $\|$ $c$). This result confirm the single band nature of the superconductivity, in agreement with ARPES data. Whereas the data can be accounted for by a negligibly small Pauli pair breaking contribution for \textbf{H} $\|$ $c$, Maki parameter $\alpha$ = 1.4, corresponding to Pauli field $\mu_0H_P$ = 79 T accounts for the data for \textbf{H} $\|$ $ab$. Coherence lengths, deduced from orbital fields, and temperature sensitivity of the critical fields for the two field directions are in good agreement with resistivity data. However, orbital fields are significantly lower than those deduced from specific heat measurements. This result suggests that, in addition to magnetic London penetration depth, other contributions enter the TDO frequency variations.

Finally, the anisotropy of the critical magnetic field, which is already rather small around $T_c$ ($\gamma$ = 1.6) becomes lower than 1 below 4 K. This result, which illustrates the commonly observed decrease of the anisotropy in iron-based superconductors as the temperature decreases, remains to be explained.

\section{Acknowledgements}

The authors wish to acknowledge fruitful discussions with Vitaly A. Gasparov. This work has been supported by EuroMagNET II under the EU Contract No. 228043 and by the French National Research Agency, Grant No. ANR-09-Blanc-0211 SupraTetrafer.


 \end{document}